\documentclass{appolb}
\usepackage{graphicx}
\usepackage{overpic}
\usepackage{hyperref}
\usepackage{cite}
\usepackage{commath}
\usepackage{authblk}
\usepackage{lineno}

\begin{document}

\title{GNN-based track reconstruction for MUonE experiment}

\author[1]{Damian Mizera}
\author[1,2]{Mateusz Goncerz}
\author[1]{Izabela Juszczak}
\author[1]{Marcin Kucharczyk
\thanks{Contact author: marcin.kucharczyk@ifj.edu.pl}
}
\author[1]{Marcin Wolter}
\author[1]{Milosz Zdybal}

\affil[1]{Institute of Nuclear Physics Polish Academy of Sciences, Krakow, Poland}
\affil[2]{AGH University of Krakow, Poland}

\maketitle

\begin{abstract}
{\bf Abstract}
A study of a Graph Neural Network–based model for track reconstruction in the context of MUonE experiment is presented, using simulated data corresponding to the test-run MUonE detector setup. The fully three-dimensional model successfully addresses both the reconstruction and particle identification challenges essential for achieving the experiment’s primary physics goal. It provides  significantly faster pattern recognition than classical reconstruction algorithms, while maintaining comparable efficiencies and resolutions.
\end{abstract}

\section{Introduction}
The measured mass of the Higgs boson and direct searches for particles beyond the Standard Model at the Large Hadron Collider (LHC)~\cite{HiggsATLAS,HiggsCMS} indicate that the energy scale for directly observing New Physics is likely much higher than previously expected. Therefore, precise indirect measurements have gained importance, as they can offer insight into theoretical extensions of the Standard Model and provide a complementary approach to LHC direct searches. One particularly promising sector for probing New Physics is the measurement of the muon anomalous magnetic moment, $a_{\mu} = (g-2)/2$. This quantity is one of the most precisely measured in experiments and most accurately predicted within the Standard Model, making it a stringent test of the theory. For the past two decades, $g-2$ experiments~\cite{g2BNL,g2Fermilab1,g2Fermilab2} have consistently indicated a significant deviation from the Standard Model prediction, with steadily improving precision on $a_{\mu}$ measurement. However, further increasing the significance of a potential discovery is hindered by theoretical uncertainties, dominated by the leading-order hadronic vacuum polarization contribution, which cannot be determined using perturbative QCD methods. To overcome this limitation, the MUonE experiment has been proposed~\cite{MUonE}. Its goal is to measure the hadronic contribution to the running of the electromagnetic coupling in the momentum-transfer region relevant for calculating the muon $g-2$ anomaly, allowing to significantly increase the sensitivity to the potential discovery of New Physics phenomena. The idea behind the MUonE experiment is to use elastic muon–electron scattering to precisely determine the hadronic contribution to $a_{\mu}$~\cite{MUEScatt}, which requires collecting large data statistics and maintaining tight control over systematic effects. The experiment will operate on the M2 muon beam at the CERN SPS~\cite{SPS}, where  high beam intensity will enable the accumulation of a vast number of events. Consequently, an efficient trigger system is essential to cope with tight timing constraints imposed by a software trigger stage running at high-frequency. This makes the development of novel techniques for effective online data reduction crucial, in order to maximize the statistical power of the final physics measurement. These techniques must allow full track and vertex reconstruction in real time within a hardware-triggerless environment. A promising alternative to conventional reconstruction procedures is the use of machine-learning methods~\cite{MLKaplan}, which have proven highly effective in pattern-recognition tasks and are widely used across many fields. Although such methods have not yet been deployed in high-energy physics experiments, they are being actively developed and are expected to be implemented in the near future. In this paper, we present studies on the development and performance of a track reconstruction approach based on Graph Neural Network (GNN) technique~\cite{GNN} designed to meet the MUonE experiment’s requirements. This method offers an optimal inductive bias, reduced number of parameters, highly refined loss function, and - crucially - a natural representation of the underlying data.

\section{MUonE experiment}
\label{sec:muone}
\begin{figure}[]
    \begin{center}
      \includegraphics[width=1.0\linewidth]{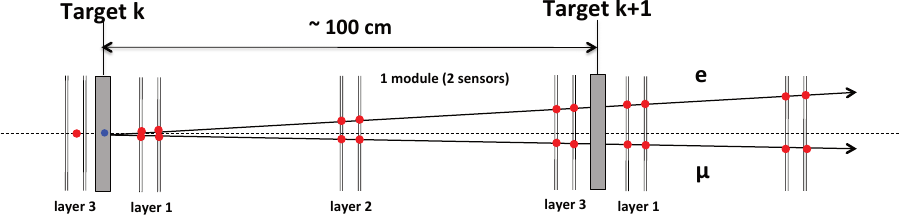}
      \vspace*{-0.5cm}
    \end{center}
    \caption{
      \small Schematic view of a single station of MUonE experiment, repeated 40 times in the final apparatus~\cite{MUonE}.
    }	
    \label{fig1}
\end{figure}

\begin{figure}[]
    \begin{center}
      \includegraphics[width=1.0\linewidth]{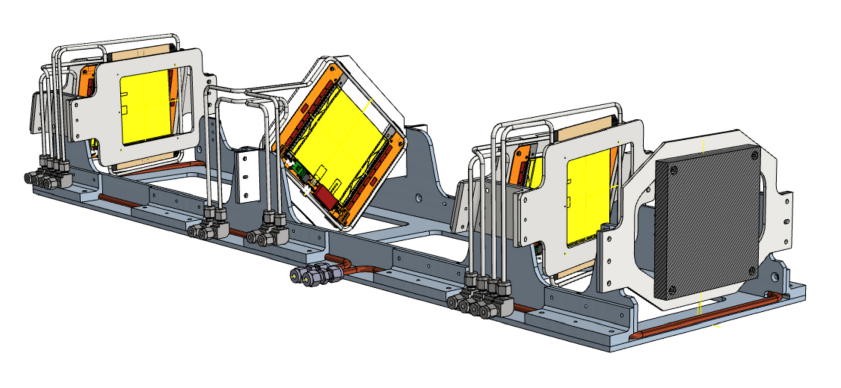}
      \vspace*{-0.5cm}
    \end{center}
    \caption{
      \small Model of a single MUonE tracking station~\cite{MUonE}.
    }	
    \label{fig2}
\end{figure}

In the MUonE experiment~\cite{MUonE}, data samples of elastic $\mu e \rightarrow \mu e$ scattering will be collected using 150–160~GeV muons incident on atomic electrons in Beryllium targets. For this purpose, the upgraded M2 muon beam at the CERN SPS~\cite{SPS} will be employed, providing high-energy, high-intensity muon and hadron beams, as well as low-intensity electron beams for calibration. As it has been measured by the NA64 experiment, the hadron contamination in the muon beam is remarkably low ($\pi/\mu < 10^{-6}$)~\cite{NA64}, and even much lower in the case of positrons. At a beam energy of about 160~GeV, the typical maximum intensity reaches 5~$\times$ 10$^7$~$muons / s$. The main tracking detector components of MUonE are shown in Figs.~\ref{fig1} and~\ref{fig2}. The tracking system will precisely measure the scattering angles of the outgoing electron and muon relative to the incoming muon direction. It will consist of 40 identical stations (Fig.~\ref{fig1}), each composed of a 3 cm thick Beryllium target followed by three tracking modules, each built from two silicon strip layers, providing a position measurements only in $(x,z)$ or $(y,z)$ plane (Fig.~\ref{fig2}). The middle module in each station is rotated by 45$^{\circ}$ around the beam axis (stereo module). The stations are spaced by roughly one meter with air gaps in between, providing a distributed low-$Z$ target together with a high-precision tracker. The state-of-the-art silicon strip sensors adopted for MUonE come from the CMS Tracker upgrade~\cite{sensorsCMS}. They offer a large active area that fully covers the MUonE acceptance and provide the required spatial resolution. Their front-end electronics support the 40~MHz readout rate needed for MUonE. The sensors are 320~$\rm{\mu}$m-thick $n$-in-$p$ devices with an active area of 10 $\times$ 10 cm$^2$. The strips are capacitively coupled with a 90~$\rm{\mu}$m pitch and segmented into two approximately 5~cm long sections. The DAQ system is also adapted from the one developed for CMS sensors for the HL-LHC upgrade. Downstream particle-identification detectors will be installed to resolve muon-electron ambiguities: an electromagnetic calorimeter for electron identification and a muon filter for muons. The calorimeter -- placed downstream of all tracking stations -- will provide particle identification, electron-energy measurement, and event selection. It will consist of lead tungstate (PbWO$_4$) crystals, similar to those used in the CMS electromagnetic calorimeter~\cite{caloCMS}, offering fast scintillation, good light yield, and compact geometry. Each crystal has a transverse size of 2.5~$\times$~2.5~cm$^2$, a length of 23~cm ($X_0$\footnote{$X_0$ is the radiation length.}=~26~g/cm$^{2}$), and will be read out with solid-state sensors (SiPMs or APDs). The calorimeter’s transverse area is about 1~$\times$~1 m$^2$, covering electrons with energy E~$\geq$~30~GeV and angular acceptance of $\theta < $~5~mrad for E~$\geq$~10 GeV. Since the muons have long lifetime and low interaction probability, they traverse the entire detector. A muon chamber will therefore be placed at the end of the setup to filter out other charged particles. The development of the MUonE detector subsystems and front-end electronics is already well advanced. They have undergone preliminary beam tests with data from 2018~\cite{testRun2018}, followed by tests with several detector modules in late 2021, and were validated during MUonE test-runs in 2022–2025~\cite{MUonEProposal}.


\section{Event reconstruction in MUonE}

\subsection{Online event reconstruction requirements}
The MUonE experiment will operate at high-intensity beams. The readout electronics must function at a frequency of 40~MHz~\cite{MUonE}. This corresponds to an average rate of about 1.25 incoming muons per clock cycle during the SPS spill. An effective online selection is therefore essential to reject the majority of uninteresting events at 40~MHz while maintaining high efficiency for genuine $\mu e \rightarrow \mu e$ scatterings. In most cases, the stations are traversed by one or two non-interacting beam muons. When interactions occur, the dominant background sources are $e^{+} e^{-}$ pair production and muon–nucleon collisions. A reduction of the event rate down to 400~kHz is required, making a fast and efficient trigger system indispensable. The MUonE trigger is foreseen to consist of two software-based levels. The first level processes signals from the silicon micro-strip sensors and runs on FPGAs, where potential track segments -- {\it stubs} -- are formed from pairs of consecutive hits within a tracking module~\cite{MUonE}. Tracks are subsequently built from such stubs, and the resulting candidates are checked against the kinematic criteria expected for $\mu e \rightarrow \mu e$ scattering. It is worth noting that requiring exactly one reconstructed incoming track eliminates any contribution from pileup effect. The second trigger level performs real-time, offline-quality event reconstruction. It must provide fast pattern-recognition algorithms for efficient track reconstruction, maintain high acceptance for $\mu e \rightarrow \mu e$ events, and ensure a sufficient reduction of the total output bandwidth. To overcome the bottleneck of limited offline storage, high-quality reconstruction and event selection must be moved from the offline to the online stage. This strategy allows the experiment to fully exploit the high-luminosity beams and deliver high-purity samples of the processes under study. All things considered, new computing techniques based on machine learning are supposed to be employed to reach the offline quality reconstruction ensuring minimal loss of signal events.

\subsection{Offline event reconstruction}
\label{offline}
Present MUonE offline algorithms~\cite{testRun2018,MUonEProposal} prioritize the efficiency of reconstructing $\mu e \rightarrow \mu e$ scattering vertices within the subset of recorded events passing the trigger. Early bias is avoided by considering every possible stub grouping, representing a valid particle track candidate. This approach is only feasible with a significant event rate reduction achieved at the trigger stage. The use of silicon strip sensors enforces a three stage reconstruction of a scattering event. Initially, two-dimensional track candidates are formed independently in the $(x,z)$ and $(y,z)$ planes, as a set of all unique stub pairs from the modules located at the beginning and the end of a tracking station. Likewise, the collection of three-dimensional track candidates is then formed by considering all unique pairs of two-dimensional track candidates in the orthogonal planes. In order to reject the purely random combinations, the compatibility of each track candidate with stubs registered in the rotated stereo planes ($u$,$v$) is evaluated. From each stereo plane, the stub closest to the track is assigned to its collection, provided the distance is smaller than 10 times the sensor resolution. If no such stubs are found, the candidate is rejected from further processing. Otherwise, the resulting collection is fitted with a $\chi^2$-like metric. The remaining candidates are ordered by their quality (based on the track fit $\chi^2$ per degree of freedom and number of stubs in the collection) and undergo a clone track\footnote{Clone tracks are multiple tracks reconstructed from hits which were deposited by a single charged particle.} removal procedure. The collection is traversed from worst to best, and then the candidates sharing stubs with those considered to be of better quality are rejected. These two stages of track reconstruction constitute the pattern recognition portion of the algorithm, which corresponds in scope to the machine learning based approach described in the present paper. Reconstructed tracks in subsequent tracking stations are combined into triplets (one incoming muon before the target and outgoing muon and electron after the target), representing $\mu e \rightarrow \mu e$ vertex candidates. The constituent tracks of each scattering vertex candidate are then refitted simultaneously, under a constraint of passing through a common position within the target's volume, and the resulting collection of scattering vertices is taken as the input for the data analysis and final physics measurement.

\section{Machine Learning and Artificial Neural Networks}
Machine learning encompasses a set of techniques by which algorithms improve themselves using data. In this process, called training, a model is created that can later be used to make predictions. The model may take one of many internal structures, one of which is the focus of this paper: the artificial neural network (ANN)~\cite{ANN}. Artificial neural networks are inspired by their biological counterparts and inherit part of the nomenclature used to describe them. A network is built with perceptrons that are connected to each other, like neurons with synapses. Every artificial neuron has a set of inputs and a single output. The returned value is determined by the activation function~\cite{activFunction}, also known as the transfer function, which takes into account the sum of weighted inputs. The network organizes the perceptrons into layers, where each is connected to all neurons in adjacent layers. This is the most common arrangement, known as a feedforward neural network, with the alternative being a recurrent neural network that allows loops in the connections. The first layer of the network is called the input layer, the last is known as the output layer, and the layers in between are hidden layers. If the network has a single hidden layer, it is called a shallow network, whereas a deep network has multiple hidden layers. The response of the perceptron to a given input, defined by the activation function, allows to determine whether the neuron should be activated, sending a signal to the neurons in the next layer. The activation function used in the present study is the rectified linear unit (ReLU)~\cite{ReLU}. It is characterized by low computational cost and behavior that more closely resembles that of a biological neuron. The prevalence of the ReLU has led to the appearance of variants of this function, such as leaky ReLU (LReLU).

The main objective of the training process is to create a generalized model capable of making predictions on previously unseen data. In the case of ANN, this entails adjusting the weight for each input in all perceptrons. There are three main approaches, and the appropriate one should be chosen according to the issue in question. In the so-called {\it supervised training}, a {\it training dataset} is used, in which a desired response (a label) is assigned to each set of inputs. During training, the model’s responses are compared with the labels using a loss function (sometimes called the cost function) that grades the responses, and the model is adjusted to better fit the desired outcomes. For training feedforward networks, the so-called {\it backpropagation algorithm}~\cite{activFunction} is widely used. A gradient of the loss function with respect to the weights of the network is calculated for each pair of inputs and labels. In each step, the weights are optimized to minimize the loss function. This is achieved with the gradient descent method or, for lower computational cost, stochastic gradient descent, in which only approximations are calculated instead of the actual gradients. In the so-called {\it unsupervised training}, the training dataset is not labeled. In this case, the neural network is expected to find patterns in the data using its internal representation. This method can be used for training for anomaly detection or clustering. The third main training method, the {\it reinforcement training}, can be applied in situations without a mathematical model of the problem, e.g. autonomous driving, swarm intelligence, or artificial intelligence in video games. In this approach, intelligent agents are scored for their actions in the provided environment, attempting to maximize the total score.

For the issues described in the present paper, supervised training is performed, as the realistic Monte Carlo simulation is available for training. There is well-defined ground-truth information regarding each event from the simulation that the model should learn to reconstruct.

\section{GNN-based track reconstruction for MUonE}
Graph Neural Networks~\cite{GNN} are a class of neural networks designed to operate on graph-structured data, where information is represented as nodes connected by edges. Unlike commonly used neural networks that process vectors or grids (e.g. images), GNNs explicitly exploit the relational structure between entities. The core idea of the GNN architecture is a message passing: each node iteratively aggregates information from its neighbors and updates its representation. After several layers, a node embedding contains information from a larger neighborhood of the graph. These learned embeddings can then be used for tasks such as node classification, link prediction, or graph classification~\cite{graphClass}. For those reasons the GNN technique seems to be an optimal solution for the pattern recognition challenges faced in the MUonE experiment.

The present study concerns a fully three-dimensional reconstruction model which addresses both the event reconstruction as well as particle identification challenges in the MUonE experiment. It is a groundbreaking extension of the previous studies~\cite{CNNMUonE,DNNMUonE}, which suffered from several limitations related to using two-dimensional models.

\subsection{GNN model}
\label{gnn}
A Monte Carlo data sample of 135K signal elastic scattering events generated at Next-to-Leading Order accuracy using MESMER event generator~\cite{Mesmer} was used for the model training and testing. It corresponds to the MUonE 2025 test-run detector configuration with two fully instrumented tracking stations (see Sec.~\ref{sec:muone}) with two 9-by-9~cm, 2~cm thick graphite targets. About 80\% of data sample is used for training, while the remaining 20\% is used as a validation set.

\begin{figure}[]
    \begin{center}
      \includegraphics[width=1.0\linewidth]{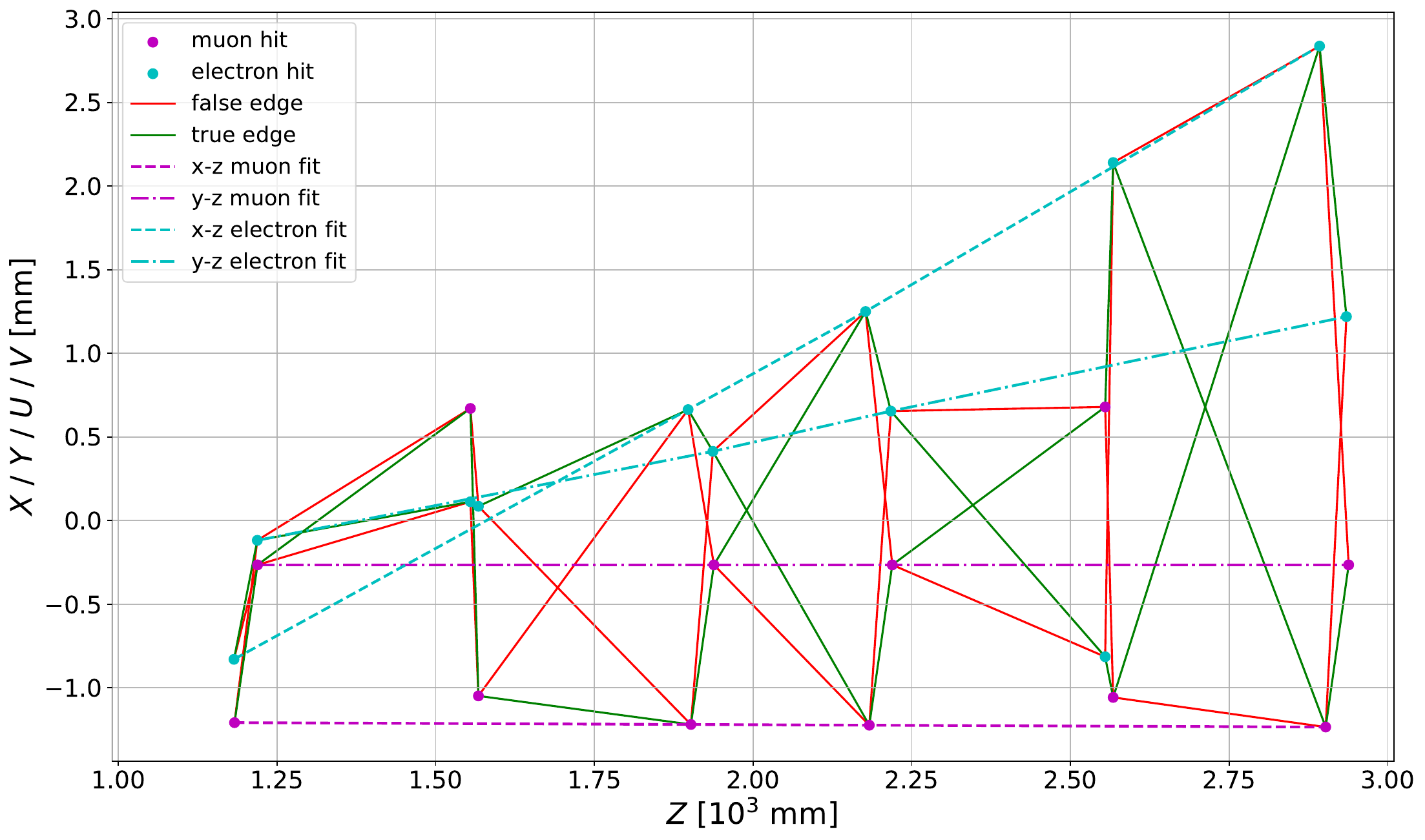}
      \vspace*{-0.5cm}
    \end{center}
    \caption{
      \small A representative training event. Edges in the resulting graph are shown as solid lines. Dashed lines connect hits from the same particle and indicate the orientation of the strip layer.
    }	
    \label{fig3}
\end{figure}

In the applied GNN-based approach, an input graph is constructed for each event, with vertices representing hit positions and edges representing all possible connections between hits in adjacent detector layers. The GNN model is trained to discriminate true edges, corresponding to the true segments of particle tracks, from the false ones that do not belong to any track. A representative event is shown in Fig.~\ref{fig3}. All hits  in detectors are marked, regardless of the layer orientation, in one plane. The horizontal axis $Z$ denotes the hit position along the beam axis, while the vertical axis represents the second coordinate of the hit position (one of $x$, $y$, $u$, or $v$), since each detector strip layer provides only two-dimensional information. Edges are formed between hits in consecutive layers independently of their orientation. Consequently, a layer orientation is not provided explicitly to the network. 

The GNN model is implemented using the~\textit{PyTorch} library~\cite{pyTorch} with the~\textit{PyTorch Geometric} extension~\cite{pyTorchGeom}, which provides tools for processing graph-structured data. The model operates on undirected graphs with features assigned to nodes only, and follows the message-passing paradigm, in which node embeddings are updated iteratively by aggregating information from neighboring nodes. During training, the network learns to assign a scalar weight to each graph edge, which can be interpreted as the probability that the corresponding connection belongs to a particle track. No explicit edge features are provided as input; instead, edge representations are learned implicitly from the features of the endpoint nodes.

The model is trained on graphics processing unit (GPU) for 500 epochs using the Adaptive Moment Estimation (\textit{Adam})  optimizer~\cite{Adam} with the {\it AMSGrad} variant~\cite{AMSGrad}. The node network employs two hidden layers with 128 neurons, while the edge network uses 256 neurons for each of the two hidden layer.  This configuration yielded satisfactory performance, however the full network optimization requires an extensive hyperparameter tuning demanding a substantial time and computational resources. Such a tuning is beyond the scope of the present work, where the main goal is to demonstrate the feasibility and potential of the proposed solution.

\subsection{Results}
\label{results}
\begin{figure}[ht]
    \centering
    \includegraphics[width=0.9\linewidth]{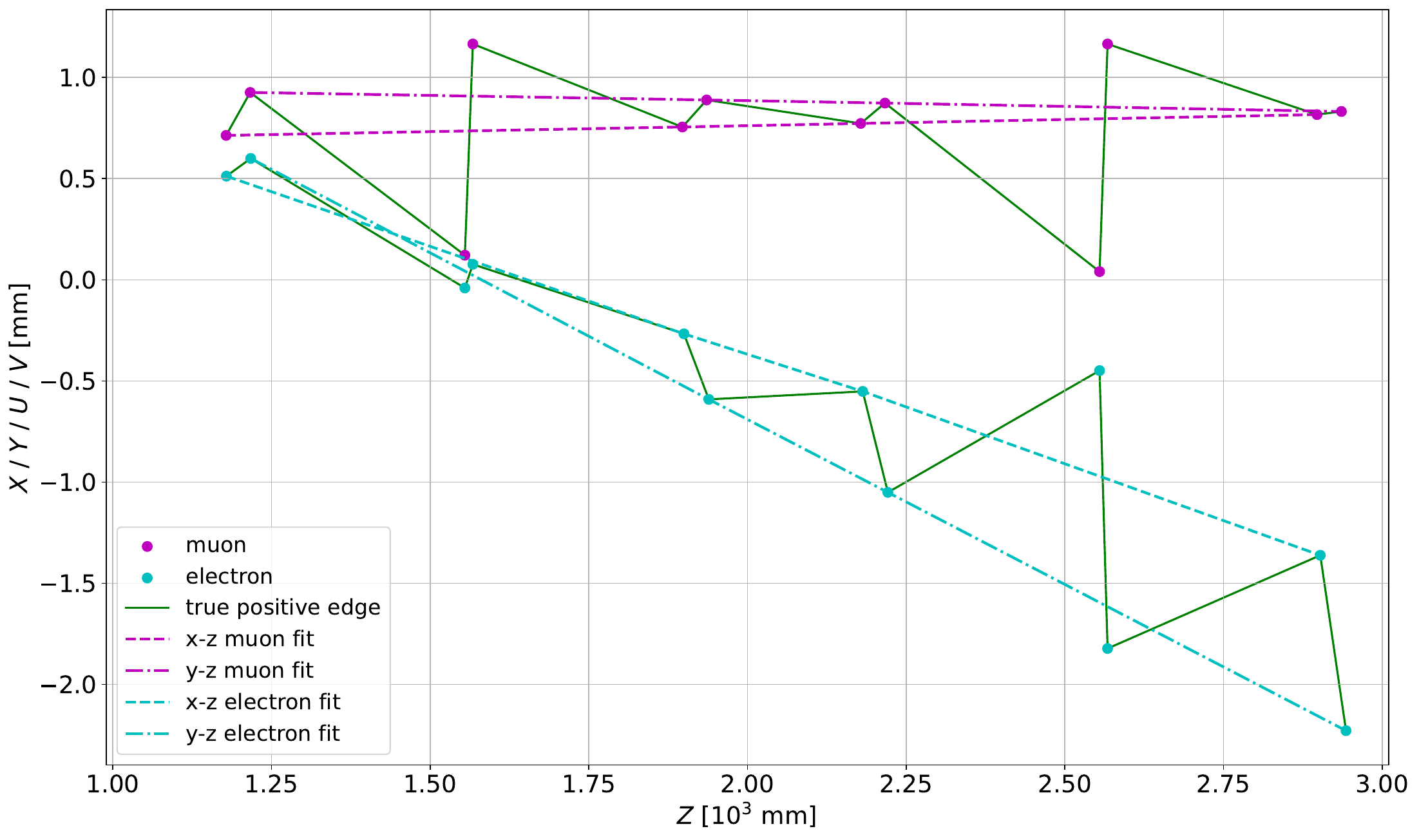}
    \caption{Signal event after edge classification. For clarity, false edges are not shown.}
    \label{fig:edges}
\end{figure}

The trained GNN model is able to efficiently recognize edges that are building the track. An example event with classified edges is shown in Fig.~\ref{fig:edges}. Hits associated with muons and electrons are displayed separately, and edges classified by the network as true track segments are indicated by solid green lines. For clarity, edges classified as false are skipped, and hits of the same type ($X$ or $Y$) from the same particle are connected by dashed lines.

\begin{figure}[ht]
    \centering
    \includegraphics[width=0.50\linewidth]{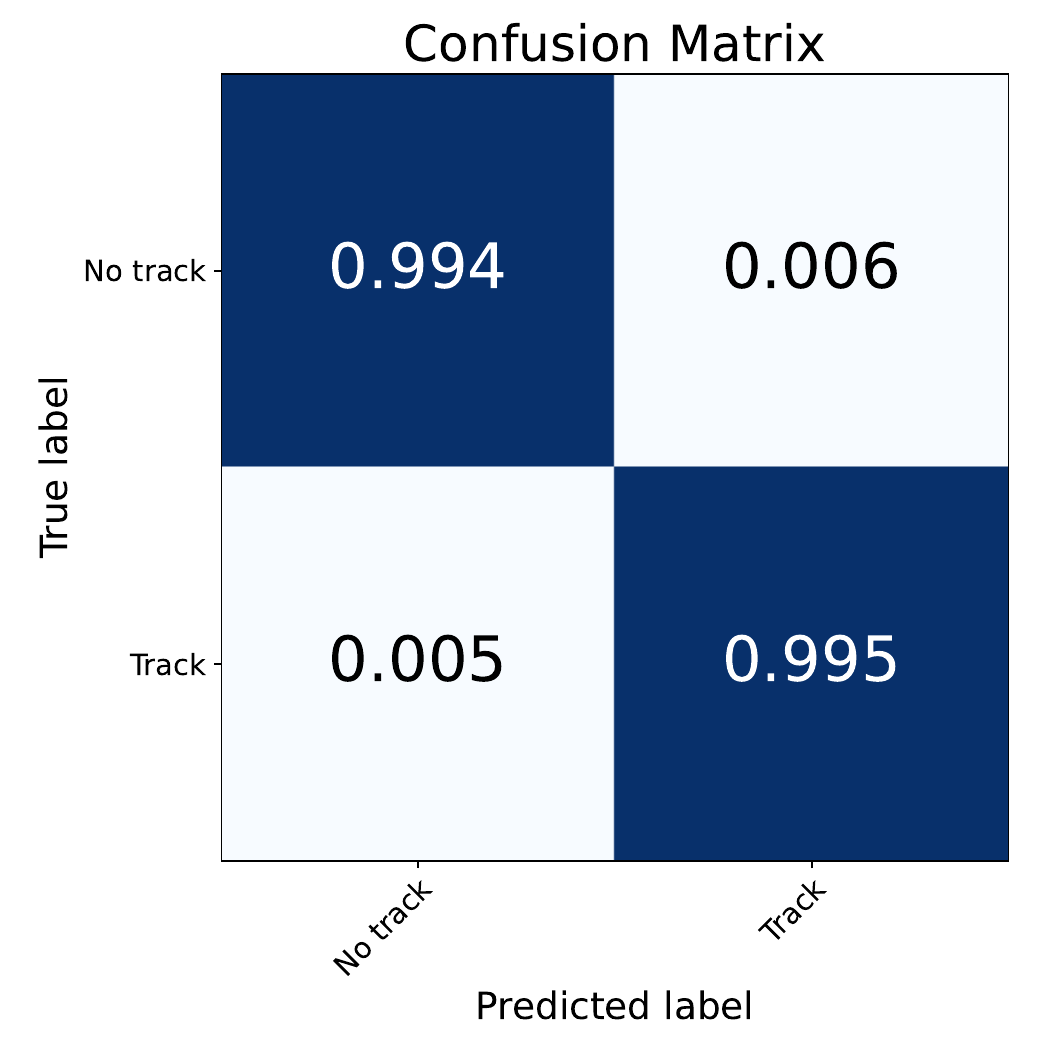}
    \caption{Confusion matrix of the network recognizing edges building a track.
    }
    \label{fig:matrixROC}
\end{figure}

The misclassification rate of the trained network does not exceed 6\textperthousand. Detailed results in the form of a confusion matrix are presented in Fig.~\ref{fig:matrixROC}.

\subsubsection{Track candidate reconstruction efficiency}
\begin{figure}[ht]
    \centering
    \includegraphics[width=1.0\linewidth]{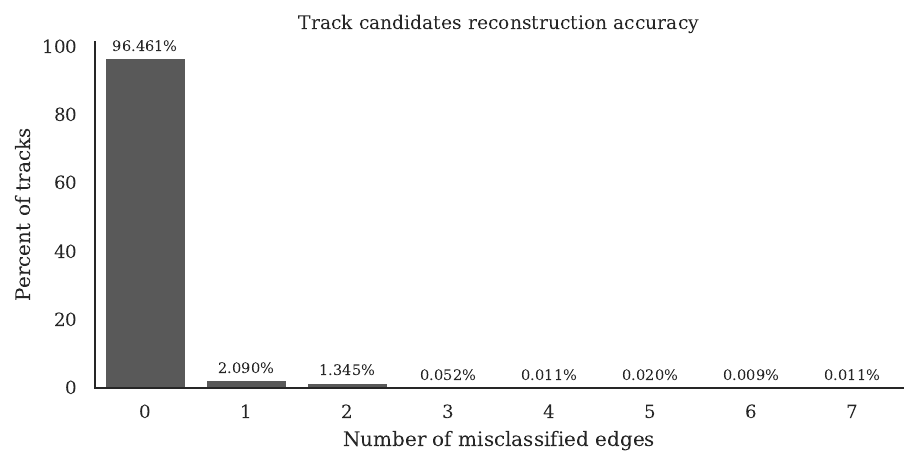}
    \caption{Number of misclassified edges in the track candidate.
    }
    \label{fig:effEdges}
\end{figure}

The efficiency to reconstruct track candidates is a key metric for pattern recognition performance. It should be emphasized that successful track reconstruction does not require perfect classification of all edges forming a track candidate, thus  much better efficiency is expected for the final track reconstruction, as a small number of misclassified or missing edges can still yield efficient reconstruction. Fig.~\ref{fig:effEdges} shows the distribution of the number of edges within a particle track that are incorrectly classified by the GNN model as not belonging to a given track. It may be observed that more than 96\% of track candidates have all edges correctly classified, while 99.9\% of tracks contain at most three misclassified edges, which still could enable proper track reconstruction.

The GNN model developed proved to maintain high track candidate reconstruction efficiency as compared with classical reconstruction algorithms, allowing to keep comparable resolutions as proved already in the previous studies~\cite{DNNMUonE}.

\subsubsection{Timing performance}
\begin{figure}
    \centering
    \includegraphics[width=0.6\linewidth]{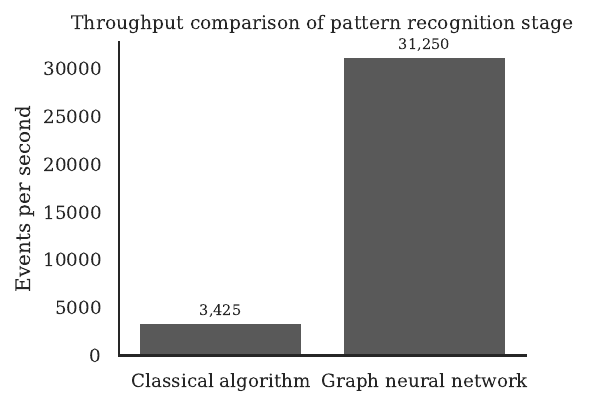}
    \caption{Throughput comparison of timing performance of the pattern recognition stage for the classical algorithm and the GNN-based method.}
    \label{fig:timing}
\end{figure}

As an efficient trigger system is essential to deal with limited CPU computing resources available online, it is crucial to implement the novel technique allowing to face with a constraint on the total processing time per event. The developed GNN model seems to be able to cope with such limitations. The processing time of the pattern recognition stage using GNN model is compared with the classical reconstruction algorithm (as described in Sec.~\ref{offline}). Both benchmarks use the same Monte Carlo signal data sample comprising over 200K graphs, and the quoted results are averaged over 30 runs. The measurements are performed on a workstation equipped with the NVIDIA RTX A5000 GPU (Ampere architecture) with 24~GB of GDDR6 ECC memory and 8192 CUDA cores, delivering a peak FP32 performance of approximately 27.8~TFLOPS, and an AMD Ryzen Threadripper PRO 5955WX CPU with 16 cores and 32 threads. In the case of GNN-based method, the pattern recognition stage encompasses graph construction, data transfer to the GPU, edge classification by the neural network, transfer of the output back to the CPU, and interpretation of the output as track candidates. The throughput comparison between timing performance for GNN-based model and classical reconstruction algorithms is shown in Fig.~\ref{fig:timing}, which clearly demonstrates almost one order of magnitude speedup for the GNN-based approach.

\subsection{Particle identification with GNN model}
\begin{figure}
    \centering
    \includegraphics[width=0.6\linewidth]{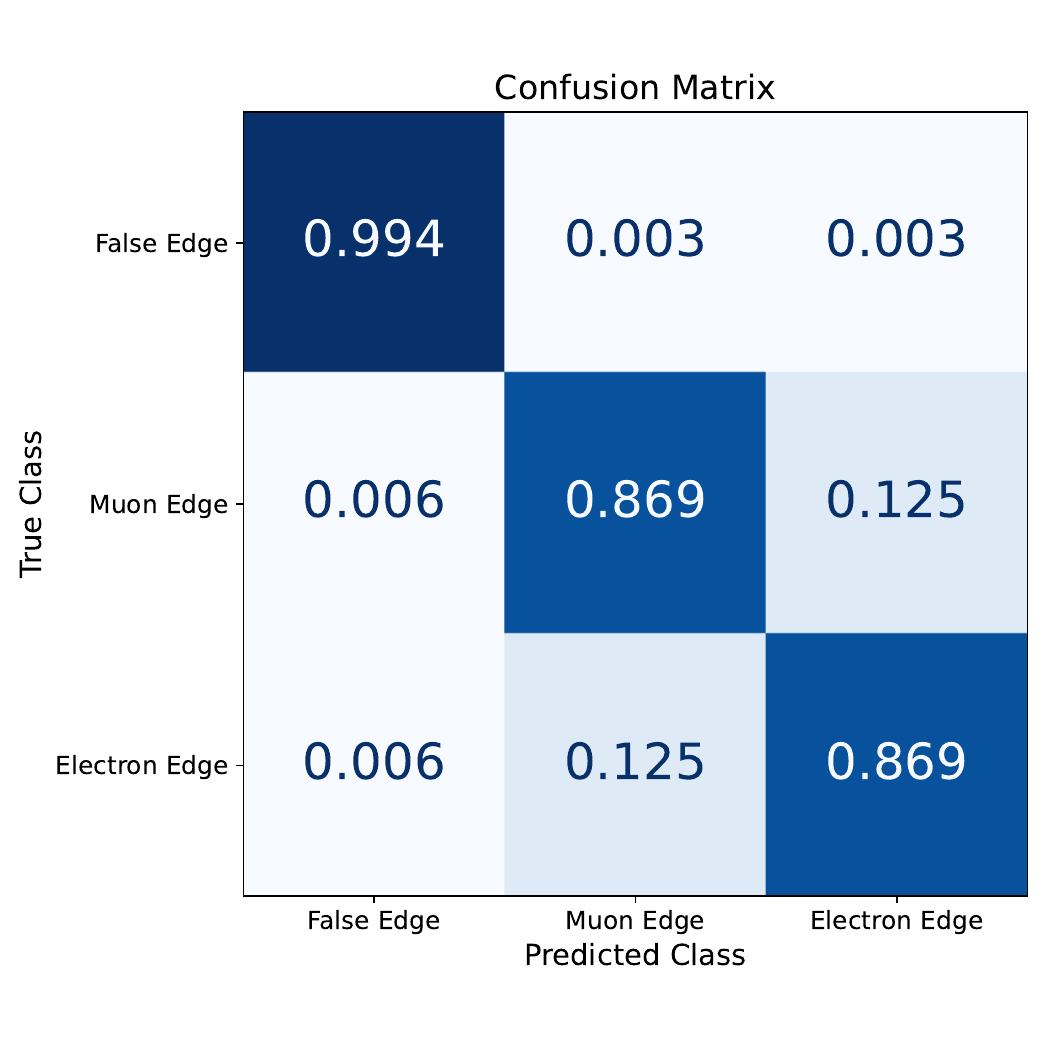}
    \caption{Confusion matrix of the network performing also particle identification.}
    \label{fig:matrixPID}
\end{figure}

\begin{figure}
    \centering
    \includegraphics[width=1.05\linewidth]{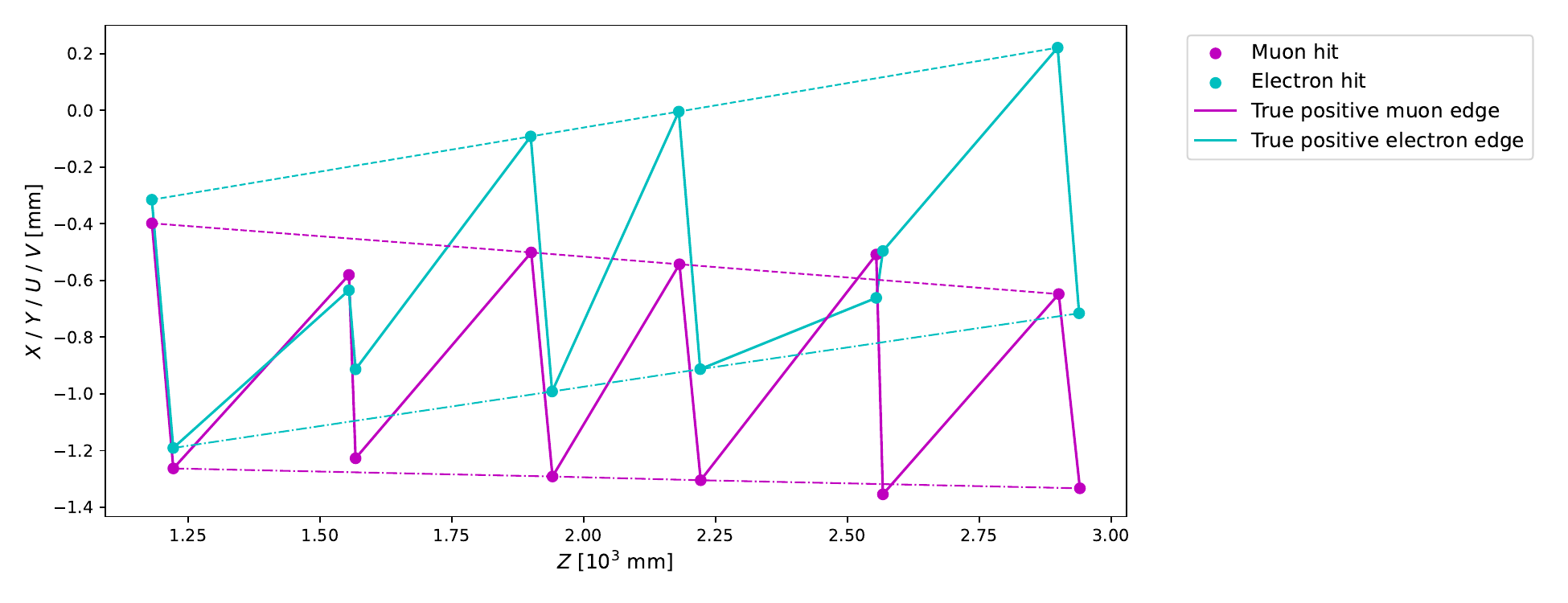}
    \caption{Example event after the GNN edge classification with particle identification. For clarity, false edges are not shown.}
    \label{fig:examplePID}
\end{figure}

With minor modifications, the GNN model described in Sec.~\ref{gnn} can be used also for particle identification (PID). Precisely, to discriminate  secondary muons from electrons simultaneously with performing a pattern recognition. In this case each graph edge is assigned a three-element label vector which components represent the probabilities that the edge is: {\it (i)} not a true track segment, {\it (ii)} true segment of muon track, or {\it (iii)} true segment of electron track. After training for 500 epochs
the model achieves similar classification performance of edges as described in subsection~\ref{results}. The accuracy of muon and  electron identification is lower and approximately 12\% of edges from muon tracks (electron tracks) are misclassified as electron (muon) edges (see Fig.~\ref{fig:matrixPID}). An example single event after GNN edge classification is shown in Figure~\ref{fig:examplePID}.

It should be noted that the primary physical observable exploited by the network to distinguish muon and electron tracks is the scattering angle in the target material. Since electrons are approximately 200 times lighter than muons, they scatter at significantly larger angles. The dominant contribution arises from scattering in the target located between the first and second detector stations, although scattering in the subsequent, thinner material between the second and third stations also contributes. 

The developed GNN-based technique for particle identification is not the main goal of the present study. However, it can serve as an alternative method with respect to classical PID algorithms, providing useful information in particular phase-space regions that may be problematic for classical particle identification as well as it may help estimating systematic uncertainty related to PID.

\section{Conclusions}
The implemented graph neural network model achieves high efficiency in classifying graph edges as potential particle track segments. Furthermore, GNN-based pattern recognition is much faster as compared to classical algorithms as it transforms a sequential, branching, combinatorially exploding optimization problem into a single, massively parallel, fixed-cost tensor inference problem executed on a GPU. Simultaneously, it satisfactorily distinguishes track segments  originating from muons and electrons. High edge classification efficiency translates into reliable full-track identification. In particular, in the MUonE experiment to reach the offline quality reconstruction at the trigger level, ensuring minimal loss of signal events, new computing techniques have to be employed. The developed GNN-based track reconstruction model can effectively cope with the CPU time limitation of standard algorithms at online stage related to growing combinatorics with increasing number of tracks (reconstructed from hits) seen in the tracking detectors. At the same time it provides a very high reconstruction efficiency. The present GNN-based model is going to be first tested on real data collected during the MUonE test-run in 2025.\\

\begin{center} {\bf Acknowledgements} \end{center}
This research was supported by the National Science Centre NCN (Poland) under the contract no. 2022/45/B/ST2/00318.

%uncomment the following lines to place a figure
%\begin{figure}[htb]
%\centerline{%
%\includegraphics[width=12.5cm]{Fig1}}
%\caption{Plot of ...}
%\label{Fig:F2H}
%\end{figure}

\end{document}